\documentclass[onecolumn]{aa}

\usepackage{graphicx}
\def\ls{{_<\atop^{\sim}}}
\def\gs{{_>\atop^{\sim}}}
\def\cgs{ ${\rm erg~cm}^{-2}~{\rm s}^{-1}$ } 

\begin{document}
%%%%%%%%%%%%%%
\title{The HELLAS2XMM survey: IV. Optical identifications and the
evolution of the accretion luminosity in the Universe.
\footnote{based on observations collected at the 
European Southern Observatory, La Silla and Paranal, Chile,
and at the Telescopio Nazionale Galileo, Roque de Los Muchachos,
La Palma, TF - Spain. Based also on observations made with XMM-Newton,
an ESA science mission. 
Table 1 is only available in electronic form 
at the CDS via anonymous ftp to cdsarc.u-strasbg.fr (130.79.128.5) 
or via http://cdsweb.u-strasbg.fr/cgi-bin/qcat?J/A+A/}}

\author{F. Fiore\inst{1}, M. Brusa\inst{2,6}, F. Cocchia\inst{1,3},
A. Baldi\inst{4,5}, N. Carangelo\inst{4}, P. Ciliegi\inst{6}, 
A. Comastri\inst{6}, F. La Franca\inst{7}, R. Maiolino\inst{8}, 
G. Matt\inst{7}, S. Molendi\inst{4}, 
M. Mignoli\inst{6}, G.C. Perola\inst{7}, P. Severgnini\inst{9}, 
C. Vignali\inst{10,6}
}

\institute {INAF-Osservatorio Astronomico di Roma \\
via Frascati 33, Monteporzio-Catone (RM), I00040 Italy.
\email{fiore@mporzio.astro.it}
\and
Dip. di Astronomia Universit\`a di Bologna
\and
Dip. di Fisica Universit\`a di Roma Tor Vergata
\and
IASF/CNR Milano
\and
Harvard-Smithsonian Center for Astrophysics
\and
INAF-Osservatorio Astronomico di Bologna
\and
Dip. di Fisica, Universit\`a Roma Tre
\and
INAF-Osservatorio Astrofisico di Arcetri
\and
INAF-Osservatorio Astronomico di Brera
\and
Dept. of Astronomy and Astrophysics Pennsylvania State University
}

\date{Received 22 April 2003 / Accepted 3 July 2003}

\abstract{ We present results from the photometric and spectroscopic
identification of 122 X-ray sources recently discovered by XMM-Newton
in the 2-10 keV band (the HELLAS2XMM 1dF sample). Their flux cover the
range $8\times10^{-15}-4\times10^{-13}$ \cgs and the total area
surveyed is 0.9 square degrees.  One of the most interesting results
(which is found also in deeper sourveys) is that about 20\% of the
hard X-ray selected sources have an X-ray to optical flux ratio (X/O)
ten times or more higher than that of optically selected AGN.  Unlike
the faint sources found in the ultra-deep Chandra and XMM-Newton
surveys, which reach X-ray (and optical) fluxes more than one order of
magnitude lower than the HELLAS2XMM survey sources, many of the
extreme X/O sources in our sample have R$\ls25$ and are therefore
accessible to optical spectroscopy.  We report the identification of
13 sources with X/O$\gs10$ (to be compared with 9 sources known from
the deeper, pencil-beam surveys). Eight of them are narrow line QSO
(seemingly the extension to very high luminosity of the type 2 Seyfert
galaxies), four are broad line QSO.  The results from our survey are
also used to make reliable predictions about the luminosity of the
sources not yet spectroscopically identified, both in our sample and
in deeper Chandra and XMM-Newton samples. We then use a combined
sample of 317 hard X-ray selected sources (HELLAS2XMM 1dF, Chandra
Deep Field North 1Msec, Chandra SSA13 and XMM-Newton Lockman Hole flux
limited samples), 221 with measured redshifts, to evaluate the
cosmological evolution of the hard X-ray source's number and
luminosity densities. Looking backward in time, the low luminosity
sources (logL$_{2-10keV}=43-44$ erg s$^{-1}$) increase in number at a
much slower rate than the very high luminosity sources
(logL$_{2-10keV}>44.5$ erg s$^{-1}$), reaching a maximum around z=1
and then levelling off beyond z=2.  This
translates into an accretion driven luminosity density which is
dominated by sources with logL$_{2-10keV}<44.5$ erg s$^{-1}$ up to at
least z=1, while the contribution of the same sources and of those
with logL$_{2-10keV}>44.5$ erg s$^{-1}$ appear, with yet rather large
uncertainties, to be comparable between z=2 and 4.


}

\authorrunning {Fiore et al.}
\titlerunning {HELLAS2XMM 1dF optical identifications}

\maketitle

%%%%%%%
\section{Introduction}

Hard X-ray observations are the most efficient way to discriminate
accretion-powered sources, such as AGN, from starlight and optically
thin hot plasma emission.  Furthermore, hard X-rays are less affected
than other bands by obscuration. The advent of imaging instruments in
the 2-10 keV band, first aboard ASCA and BeppoSAX (e.g. Ueda et
al. 1999, Akiyama et al. 2000, Della Ceca et al. 1999, Fiore et
al. 1999,2001, La Franca et al. 2002, Giommi et al. 2000) and then on
Chandra and XMM-Newton, has led to a dramatic advance.  Deep surveys
have resolved 80-90\% of the 2-10 keV Cosmic X-ray Background (CXB,
see e.g. Mushotzky et al. 2000, Giacconi et al. 2001,2002, Brandt et
al. 2001, Hasinger et al. 2001), and the detailed study of the cosmic
evolution of the hard X-ray source population is being pursued
combining deep and shallow Chandra and XMM-Newton surveys. These
studies confirm, at least qualitatively, the predictions of standard
AGN synthesis models for the CXB (e.g. Setti \& Woltjer 1989, Comastri
et al. 1995, 2001): the 2-10 keV CXB is mostly made by the
superposition of obscured and unobscured AGN. Quantitatively, though,
rather surprising results are emerging: a rather narrow peak in the
range z=0.7-1 is present in the redshift distributions from ultra-deep
Chandra and XMM-Newton pencil-beam surveys (e.g. Barger et
al. 2001,2002, Cowie et al. 2003, Hasinger 2003), in contrast to the
broader maximum observed in previous shallower soft X-ray surveys
(e.g. ROSAT, Schmidt et al. 1998, Lehmann et al. 2001) and predicted
by the above mentioned synthesis models; furthermore, evidence is
emerging (related to the difference above) of a luminosity dependence
in the number density evolution of both soft and hard X-ray selected
AGN (Cowie et al. 2003, Hasinger 2003).

The ultra-deep Chandra and XMM-Newton surveys of the Chandra Deep
Field North (1Msec, CDFN, Brandt et al. 2001), Chandra Deep Field
South (CDFS, Giacconi et al. 2002) and Lockman Hole (LH, Hasinger et
al. 2001) cover each about 0.05-0.1 square degrees. For this reason
the number of high luminosity sources in these surveys is small (the
slope of the AGN luminosity function at high luminosities is very
steep). As an example, in the CDFN there are only 6 AGN with
logL$_{2-10keV}\gs44$ at z$>3$ and 20 at z$>2$ (Cowie et al. 2003).
To compute an accurate luminosity function on wide luminosity and
redshift intervals, and to find sizeable samples of of ``rare''
objects, such as high luminosity, highly obscured type 2 QSO (and also
of other rare sources like the X-ray bright, optically normal
galaxies, XBONGs, Fiore et al. 2000, Comastri et al. 2002) a much
wider area needs to be covered, of the order of a few square degrees.
To this purpose we are carring out the HELLAS2XMM serendipitous
survey, using suitable XMM-Newton archive observations (Baldi et
al. 2002). The HELLAS2XMM survey goal is to cover $\sim4$ square
degrees using 20 XMM-Newton fields. The majority of them share a
Chandra coverage, which, thanks to the higher angular resolution,
helps in the process of optical identification (see Brusa et
al. 2003). The survey will consist of about 500 sources selected in
the 2-10 keV band. Among them, we expect at least 100 highly obscured
AGN, a number sufficient to evaluate their luminosity function and
evolution up to z=2-3 with a level of statistical accuracy adequate
for a meaningflul comparison with those of the unobscured AGN.

As an intermediate step, in this paper we present the results of the
optical identification of 122 hard X-ray selected sources detected in
five XMM-Newton fields (PKS0312-77, PKS0537-28, A2690, G158-100,
Markarian 509), covering a total of 0.9 square degrees (hence the acronym
`1dF' for this sample), and of 0.4 square degree at
F$_{2-10keV}=2\times10^{-14}$ \cgs, where $\sim40\%$ of the CXB is
resolved in discrete sources). The optical photometric and
spectroscopic observations of the sources in the other 20 fields is in
progress.  The five XMM-Newton fields 
have been studied using the ESO 3.6m and VLT/UT1 telescopes
and the TNG telescope. For all the 122 X-ray sources we have
complete photometric coverage down to R$\sim25$ and nearly complete
spectroscopic coverage down to R=24 (90\%).  We combine the results on
the optical identification of the HELLAS2XMM 1dF sources with those
from the CDFN, Chandra SSA13 (Mushotzky et al. 2000, Barger
et al. 2001), and XMM-Newton Lockman Hole fields to
evaluate the evolution of the hard X-ray selected AGN number and
luminosity density up to z=2--3.  The combined sample consists of 317
hard X-ray selected sources spanning the flux range
$10^{-15}-4\times10^{-13}$ erg cm$^{-2}$ s$^{-1}$.  The fraction of
sources with measured redshift is 70\%.

The paper is organized as follows: Sects. 2 and 3 present the
results of the optical photometric and spectroscopic identifications
of the HELLAS2XMM 1dF source sample; Sect. 4 compares the HELLAS2XMM
1dF sample with other samples of hard X-ray sources from deeper
Chandra and XMM-Newton survey; Sect. 5 discusses the population of sources
with high X-ray to optical flux ratio.  We show here how the results
obtained using the HELLAS2XMM 1dF sample can be used to make
statistical predictions about this population of sources at fainter
X-ray (and optical) fluxes; Sect. 6 presents our results on the
evolution of the X-ray source's number and luminosity density; finally
Sect. 7 discusses our main findings.  A $H_0=70$ km s$^{-1}$
Mpc$^{-1}$, $\Omega_M$=0.3, $\Omega_{\Lambda}=0.7$ cosmology is
adopted throughout.

\section{Optical identifications}

We have obtained relatively deep (R=24-25) optical images for all 122
hard X-ray sources in the HELLAS2XMM 1dF sample using EFOSC2 at the
ESO 3.6m telescope and DOLORES at the TNG. Exposures were typically of
10m per image. Optical frames and X-ray images were brought to a
common astrometric reference frame using bright AGN (from 5 to 15 AGN
per field). Typical systematic shifts were of the order of 1$''$, the
maximum shift was of $\sim2''$.

Images were bias subtracted, flat field divided, and flux calibrated
using observations of standard stars aquired during each night.
Source detection was performed using the SExtractor package (Bertin et
al. 2000).  Additional 3$''$ diameter aperture photometry was performed
on the R band images at the position of eight bright K band sources
(Mignoli et al. 2003).

\begin{figure}
\centering
\includegraphics[angle=0,width=8cm]{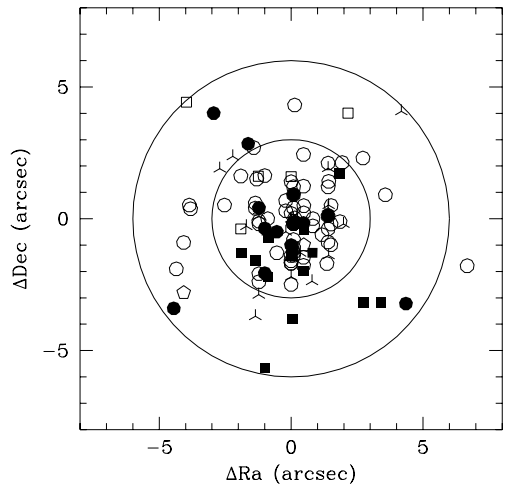}
\caption{
The displacement between the X-ray position and the position of the
nearest optical counterpart for the HELLAS2XMM 1dF sample sources. 
Open circles = broad line AGN; filled circles = narrow line AGN;
filled squares = emission line galaxies; open squares = normal galaxies;
stars = stars; pentagons = clusters of galaxies; 
skeleton triangles = unidentified objects. The two circles have radii
of 3 and 6 arcseconds.
}
\label{radec}
\end{figure}

We found optical counterparts brighter than R$\sim25$ within $\sim6''$
from the X-ray position in 116 cases (actually within 3$''$ for
$\sim80$\% of the cases, see Fig. \ref{radec}).  The average
displacement between the X-ray positions and the positions of the
optical counterparts is of $2.1''\pm1.5''$.  This is consistent with
what found in other XMM-Newton surveys (Barcons et al. 2002, Hasinger et
al. 2001).  Six X-ray sources have optical counterparts fainter than
R$\sim24$ (5\%).

Two of the sources with optical counterpart (PKS0312\_8 and A2690\_13)
are more extended than the XMM-Newton/EPIC Point Spread Function at their
off-axis angles (Ghizzardi 2001). Their X-ray emission is
probably due to intracluster or intragroup hot, optically thin, plasma
emission. For another source (PKS0537\_37) several relatively bright
galaxies are present around the X-ray error-box.  Also in this case the
X-ray emission is probably due to intracluster or intragroup
gas. These three sources will therefore be excluded when studying the
evolution of the accretion luminosity

Table 1 gives for each source the X-ray position, the position of the
optical counterpart, the displacement between the X-ray and optical
positions, the X-ray flux, the R magnitude of the optical counterpart
(or the three $\sigma$ upper limit), the classification of the optical
spectrum, the redshift and the X-ray luminosity.

\section{Optical spectroscopic redshifts and classification}

Optical spectra of 97 of the 110 sources with optical counterparts
brighter than R=24 have been obtained using EFOSC2 @ the ESO 3.6m
telescope and FORS1 @ VLT/UT1 during 5 observing runs performed
between Jan. 2000 and Aug. 2002.  A total of 9 nights at the 3.6m and
24 hours of VLT time have been devoted to this program.
\footnote{six sources have counterparts with R between 24 and 25,
still accessible to VLT spectroscopy, but demanding exposure times
too long for the time allocated to our VLT run}

Long slit spectroscopy has been carried out in the 3800-10000 \AA\ band
with resolution between 7 and 13 \AA. Data reduction was performed
using both the MIDAS (Banse et al. 1983) and IRAF\footnote{IRAF is
distributed by the National Optical Astronomy Observatories, which is
operated by the Association of Universities for Research in Astronomy,
Inc., under cooperative agreement with the National Science Fundation}
packages.  Wavelength calibrations were carried out by comparison with
exposures of He-Ar, He, Ar and Ne lamps. The flux calibration of the
spectra was obtained using observations of spectro-photometric
standard stars (Oke, 1990), performed within a few hours from the
object's spectroscopy.

For 93 sources the optical spectroscopy produced reliable redshifts.
In the four remaining cases the redshift determination is tentative,
based on a faint, single line.

%The distributions of the X-ray and optical flux of the remaining 13
%sources with optical counterpart brighter than R=24 but without
%optical spectroscopy is consistent with that of the sources with
%optical spectroscopy. Therefore the optical spectroscopic
%identifications can be considered representative of the HELLAS2XMM 1dF
%sample of sources with counterparts brighter than R=24.

\subsection{Source breakdown}

In the majority of the cases optical spectra are of sufficiently good
quality to allow a reliable classification of the optical counterpart.
Objects with permitted emission lines broader than 2000 km/s (FWHM)
are identified with type 1 AGN; objects with permitted emission lines
narrower than 2000 km/s are classified as type 2 AGN, if they possess
strong [OIII], NeV, MgII, or CIV emission lines, or as Emission Line
Galaxies (ELGs) if they possess strong [OII] or H$\alpha$ emission
lines. In six cases the presence of broad emission lines in the
optical spectrum cannot be excluded, due to the insufficient quality
of the spectrum and the classification of the object is therefore
uncertain (see notes in Table 1).  Objects without strong emission
lines (equivalent width EW$<5-8$ \AA\ )but with stellar absorption lines
and a red continuum are classified as Early Type Galaxies (ETGs).

In four cases there are two candidate counterparts inside the X-ray
error-box (see notes in Table 1) which could both contribute to the
detected X-ray emission. In two cases the counterparts are at the same
redshift and are both emission line galaxies, hence the X-ray
luminosity is univocally determined, of course within a factor of two.
In one case the two counterparts are a galaxy and a type 1 AGN,
which is therefore adopted as the most likely counterpart. The last
case, PKS0312\_35, is rather intriguing: both a type 1 QSO and an
extremely red object, R-K$>6.6$, which is also a 0.4mJy (at 5GHz)
radio source, see Brusa et al. (2003), are present in the
error-box. In the following we shall adopt as counterpart the type 1
QSO.

The source breakdown therefore includes: 61 broad line QSO and Seyfert
1 galaxies; 14 narrow line AGN (9 of which have log$L_{2-10keV}>44$
erg s$^{-1}$, see next Sect. for details on luminosity
determination, and can therefore be considered type 2 QSO); 14
emission line galaxies, all with log$L_{2-10keV}>42.7$ and therefore
all probably hosting an AGN; 5 early type galaxies with
$41.9<$log$L_{2-10keV}<43.0$, therefore XBONGs, all probably hosting
an AGN; 1 star; 2 groups or clusters of galaxies.

In summary, 94 of the 97 sources with optical spectroscopy are
associated with AGN emission, the majority (63\%)
are type 1 AGN.

A more detailed discussion of the optical spectra of all objects is
deferred to another publication.  In the following we limit ourselves
to consider two broad categories: {\em optically unobscured AGN},
i.e. type 1, broad emission line AGN, and {\em optically obscured
AGN}, i.e. AGN whose nuclear optical emission, is totally or strongly
reduced by dust and gas in the nuclear region and/or in the host
galaxy.

\section{The HELLAS2XMM 1dF sample in a context: comparison
with deeper Chandra and XMM-Newton surveys}.

In the following Sects. we shall also use results from three
additional hard X-ray samples:

1- Chandra Deep Field North sample from Barger et al. (2002):
88 sources with F$_{2-10keV}>10^{-15}$ \cgs in the inner 6.5 arcmin
radius region plus 32 sources with F$_{2-10keV}>5\times10^{-15}$ \cgs
in the annulus between 6.5 and 10 arcmin radii, for a total of 120 sources
(67 with a spectroscopic redshift).

2- Lockman Hole sample: 55 sources in the inner 12 arcmin radius
region. Fluxes are from Baldi et al. (2002), optical identifications are
from Mainieri et al. (2002), for a total of 55 sources, 44 already
identified (41 spectroscopic redshifts and 3 photometric redshifts);

3- SSA13 sample from Mushotzky et al. (2000) and Barger et
al (2001): 20 sources with F$_{2-10keV}>3.8\times10^{-15}$ \cgs
in a 4.5 arcmin radius region, 13 with spectroscopic redshift.
 
Overall, we shall deal with 317 hard X-ray selected sources from
Chandra and XMM-Newton surveys, 221 (70\%) of them identified with an optical
counterpart whose redshift is available.  Five of these sources are
identified with groups or cluster of galaxies, two with stars.  We
shall not consider in the following these seven sources.  It is highly
unlikely that other stars are present among the sources without a
redshift.  It is also unlikely that other groups or clusters of
galaxies are present among the sources without an optical counterpart
(nine in the whole combined sample), because the X-ray sources do not
present any evidence of extension. In the following we shall assume
that no other star, group or cluster of galaxy is present among the
sources without a redshift and limit the analysis to the 310 sources
probably hosting an active nucleus.

Classification of the optical spectra from the additional samples are
adopted from the above publications, when available. Otherwise, we
produced a tentative classification by visual inspection of the
published optical spectra (Barger et al. 2001,2002), adopting the
criteria in Sect. 3.1.

The combined sample includes 113 broad line AGN and 108 optically
obscured AGN. 

\subsection{Luminosities}

A key ingredient to estimate rest frame 2-10 keV, absorption
corrected, luminosities is the shape of the X-ray spectrum.  The
discussion of the detailed X-ray properties of the HELLAS2XMM sources
is beyond the purposes of this paper and will be presented in future
publications (Baldi et al. in preparation, Perola et al. in
preparation). Nevertheless, a softness ratio, defined as (S-H)/(S+H),
where S and H are the 0.5-2 keV and 2-10 keV fluxes respectively, can
be used to derive reliable assumptions about the spectral shape.
Lower values of the softness ratio indicate hard spectra; note that a
value of -1 is by definition an absolute minimum of this quantity.

\begin{figure}
\centering 
\includegraphics[angle=0,width=9cm]{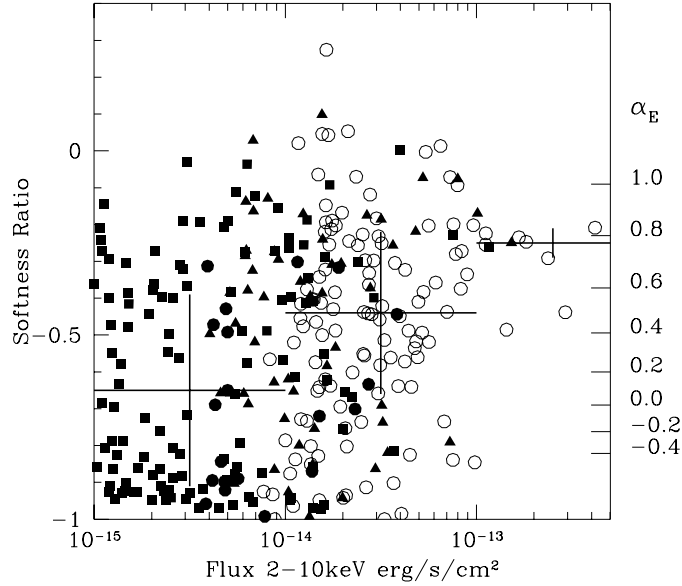}
\caption{
The softness ratio (S-H)/(S+H) (S=0.5-2 keV flux; H=2-10 keV flux)
as a function of the 2-10keV flux
for the sources in the combined sample
(HELLAS2XMM = open circles; CDFN = filled squares; LH = filled triangles;
SSA13 = filled circles).
The thick crosses represent the median with the quartile range,
in three flux intervals. The right axis gives the conversion from
(S-H)/(S+H) to $\alpha_E$ for a single power law spectrum.
}
\label{hrt}
\end{figure}

Fig. \ref{hrt} shows (S-H)/(S+H) as a function of the 2-10keV flux
for the combined sample.  At all fluxes from $10^{-15}$ to $10^{-13}$
\cgs (S-H)/(S+H) appears to span a very large range of values, which
could hardly be accounted for by statistical uncertainties on each
value, although this in some cases can be as large as 0.4-0.5.
Furthermore, the thick crosses, which represent the median with the
quartile range in three flux intervals, show a general trend
in the spectral shape, which becomes harder as the flux decreases (see
e.g. Giacconi et al. 2002 and references therein).  If the spectrum is
parameterized as a single power law with energy index $\alpha_E$
($F(E)\propto E^{-\alpha_E}$) it is straightforward to convert
(S-H)/(S+H) into the slope $\alpha_E$, and indicative values are given
on the right axis in Fig. \ref{hrt}.  We note that this is not a
completely self-consistent procedure, because fluxes in the two bands
were obtained assuming a given spectral shape (Baldi et al. 2002).
However, because the width of the energy bands is small, $\ls0.7$
decades, the conversion factors between count rates and fluxes depend
little on the spectral shape assumption. For example the conversion factors
change by 15\% and 6\% for a $\Delta\alpha=0.5$ in the 2-10 keV and
0.5-2 keV bands respectively.  This is much smaller that the typical
statistical errors on (S-H)/(S+H) and than the spread in (S-H)/(S+H)
in Fig. \ref{hrt}.  If the single power law is the correct model,
the points would span a very large interval in spectral slopes, from
about 1.3 to -0.4 or even harder. Such a distribution is not
observed in any sample of AGN, for which $\alpha_E$ (in the observed
0.5-10 keV band) is hardly ever outside the 1.3-0.5 range.  On the
other hand, photoelectric absorption would easily produce the observed
softness ratios (a column density of $10^{23}$ cm$^{-2}$, at z=0,
would give (S-H)/(S+H) lower than -0.9 for $\alpha_E=0.8$), and it is
observed in the spectra of a large fraction of bright
AGN. Furthermore, the hardening of the sources toward lower fluxes is
consistent with the expectation of AGN synthesis models, which
reproduce the CXB through the superposition of
obscured and unobscured AGN (see e.g. Comastri et al. 2001).  Our
conclusion is that the wide distribution in softness ratio is unlikely
to be solely due to an intrinsic dispersion of the spectral index, and
that the presence of substantial absorbing columns is likely the
dominant cause of its extension, expecially toward values less than,
say, --0.5.  We have therefore decided to adopt a power law reduced at
low energy by photoelectric absorption as the model used to compute the
K-correction, according to the following procedure.

We assumed a power law index $\alpha_E=0.8$ and used the observed
softness ratios to estimate the column density of the absorber and
therefore the unabsorbed fluxes. These unabsorbed fluxes were then
used to compute rest frame 2-10 keV luminosities.  The correction due
to the photoelectric absorption is small in most cases ($\ls10-20\%$
for 80-90\% of the sources). For the rest of the sources having
softness ratios suggesting observer frame column densities of the
order of $\gs3-5\times10^{22}$ cm$^{-2}$, the correction can be of the
order of a few. In a few cases the observed flux could be due to
reflection, rather than to direct emission from the AGN. In these
cases the real luminosities could be much higher than our estimates
based on the very simple spectral model adopted.  In many cases, the
uncertainty on the correction can be of the same order of the
correction itself, given the large errors on the determination of the
softness ratio and therefore of the absorbing columns.  However, since
in sources with measured redshift we find no correlation between the
softness ratio, and therefore the observer frame column density, and
the redshift, this problem is not likely to introduce a bias in the
evaluation of the evolution of the AGN luminosity function and its
integrals.

On the other hand, the luminosity of all high redshift sources depends
critically on the assumed X-ray spectral index.  Changing the spectral
index by $\Delta\alpha=0.5$ would imply a change in luminosity of a
factor of 2 at z=3. We shall return on this point in Sect. 6.

\begin{figure}
\centering 
\includegraphics[angle=0,width=9cm]{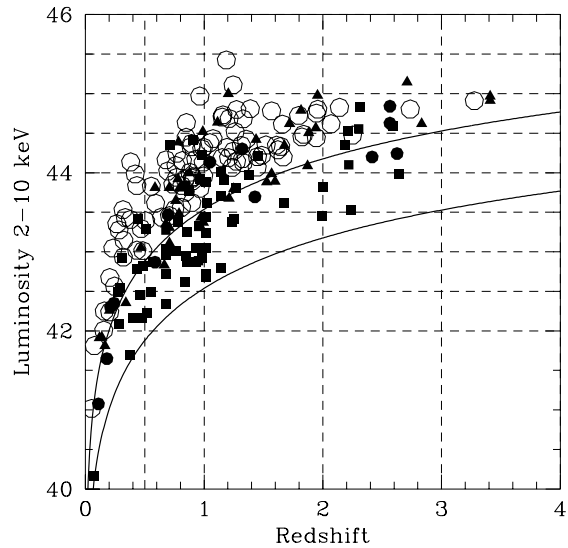}
\caption{
The 2-10keV luminosity as a function of the redshift for the
sources in the combined sample. Symbols as in figure \ref{hrt}.
The lower and upper solid lines represent the flux limits
of $10^{-15}$ \cgs (i.e. Chandra deep surveys) and $10^{-14}$ \cgs (i.e.
HELLAS2XMM 1dF survey) respectively. 
}
\label{zlx}
\end{figure}

The redshift-luminosity diagram for the sources in the combined sample
is shown in Fig. \ref{zlx}. It illustrates the fact that narrow beam
surveys at z$<1-2$ can hardly find objects of L$_X\gs10^{44}$ erg
s$^{-1}$, i.e. the sources closer to the break in the Luminosity
Function, which contribute most to the luminosity density at that
redshift. Hence the value added by larger area surveys, such as the
HELLAS2XMM survey.

\section{The X-ray to optical flux ratio: the extreme X/O source population}

\begin{figure*}
\centering
\includegraphics[angle=-90,width=15cm]{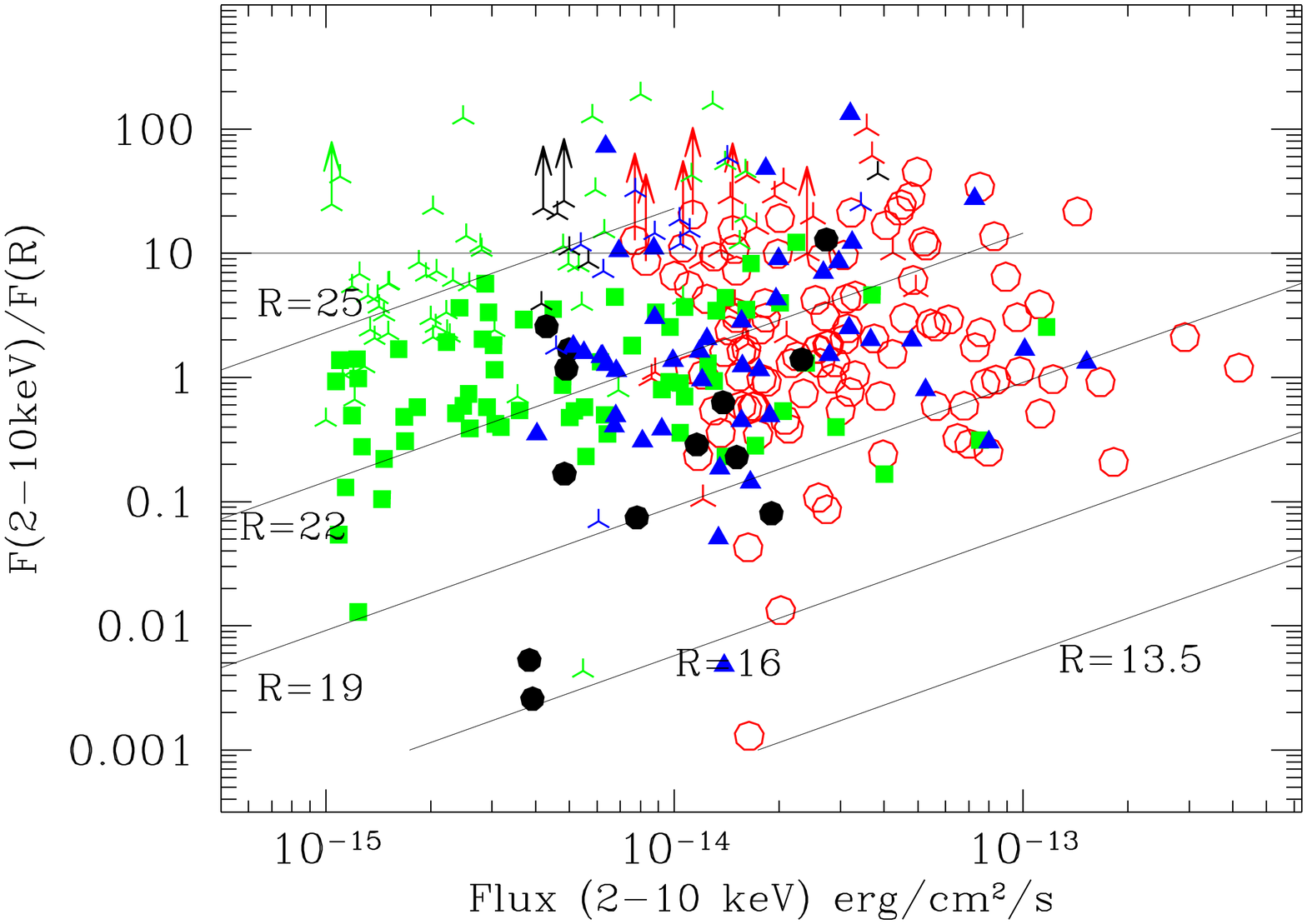}
\caption{ The X-ray (2-10 keV) to optical (R band) flux ratio X/O as a
function of the X-ray flux for the combined sample (symbols as in
Fig. \ref{hrt}, skeleton triangles are sources without a measured
redshift).  Solid lines mark loci of constant R band magnitude.  The
part of the diagram below the R=25 line is accessible to optical
spectroscopy with 10m class telescopes. Note as $\sim20\%$ of the
sources have X/O$\gs10$, irrespective of the X-ray flux.  HELLAS2XMM
1dF sources with X/O$\gs10$ have R=24-25, and therefore their
redshifts can be measured through optical spectroscopy.  }
\label{xo}
\end{figure*}

Fig. \ref{xo} shows the X-ray (2-10 keV) to optical (R band) flux
\footnote{The R band flux is computed by converting R magnitudes in
specific fluxes and then multiplying by the width of the R filter.  We
used $f_R(0)=1.74\times10^{-9}$ erg cm$^{-2}$ s$^{-1}$ $\AA^{-1}$ and
$\Delta\lambda_R=2200 \AA$, Zombeck (1990).}  ratio (X/O) as a function
of the X-ray flux for the four surveys. The part of the diagram below
the R=25 line is accessible to optical spectroscopy with 10m class
telescopes. On the other hand, the redshift of objects with
$25\ls$R$\ls27$ could be determined through the comparison of their
optical-NIR photometric spectral energy distribution 
to galaxy and AGN templates.  This is also
feasible with today 10m class telescopes.  Redshift estimate of
even fainter sources must await next generation facilities like
the 30-50m ground based telescopes and NGST, unless this can be done
using directly the X-ray spectrum, i.e.  by detecting and identifying
narrow emission or absorption features.

The range of X/O spanned by X-ray sources is extremely large, up to 6
dex or even more. Indeed, one of the most intriguing results of hard
X-ray surveys, is that about 20\% of the sources has an X-ray to
optical ratio 10 times higher or more than that typical of broad line
AGN (X/O$>10$ in Fig. \ref{xo}, while typical PG QSO and Seyfert 1
galaxies have X/O$\sim1$, see also Alexander et al. 2001).  The ratio
between the optical to X-ray optical depth, in the observer frame,
scales roughly as $(1+z)^{3.6}$, because dust extinction increases in
the UV while X-ray absorption strongly decreases going toward the high
energies.  The net result is that in the presence of an absorbing
screen the observed optical flux of high-z QSO can be strongly
reduced, and the observed magnitudes can be mainly due to starlight in
the host galaxies. Conversely, the 2-10 keV X-ray flux can be much
less reduced.  Many extreme X-ray to optical ratio sources could then
be distant, highly obscured QSO, i.e. type 2 QSO.

At the flux limit of the HELLAS2XMM 1dF survey
(F$_{2-10keV}\sim10^{-14}$ \cgs) the depth of the optical photometry
and spectroscopy allows the identification of sources with
X/O$\sim10$.  Conversely, most of the X/O$\gs10$ objects from
ultradeep Chandra and XMM-Newton surveys (e.g. CDFN, CDFS) have R$>25$
and therefore are not accessible to optical spectroscopy and require
ultradeep optical photometry.  The paucity of type 2 QSO among the
identified sources in these surveys might therefore be partly due to a
selection effect.

Among the 310 sources from the three samples considered here there are
65 sources with X/O$>10$ (21\%), 28 of them come from the HELLAS2XMM
1dF survey, and we have obtained reliable VLT spectra of 13 of them
(46\%).  This has to be compared with the 9 out of 38 identifications
in the CDFN, SSA13 and LH samples (including 3 photometric redshift of
faint optical sources in the LH sample).  Among the 13 identified
X/O$>10$ HELLAS2XMM sources we find 8 type 2 QSO, based on the absence
of broad optical lines and their high X-ray luminosity
(L$_{2-10keV}>10^{44}$ erg s$^{-1}$).  We also find 4 broad line
QSO. The remaining source has an optical spectrum typical of a narrow
emission line galaxy and has an X-ray luminosity of
L$_{2-10keV}=7\times10^{43}$ erg s$^{-1}$.

Table 2 gives the total number of sources and the number of
spectroscopic identification in two flux
intervals (F$_{2-10keV}=10^{-15}-10^{-14}$ \cgs and
F$_{2-10keV}\gs10^{-14}$ \cgs) and in three bins of X/O 
($>10$; 3--10; 0.1-3. 11 sources have X/O$<0.1$).

\setcounter{table}{1}    

\begin{table*}[ht]
\caption{\bf Number of sources per bins of flux and X/O ratio}
\begin{tabular}{lcccccc}
\hline
\hline
\multicolumn{1}{l}{ }&
\multicolumn{3}{l}{F$_{2-10keV}<10^{-14}$ \cgs}&
\multicolumn{3}{l}{F$_{2-10keV}>10^{-14}$ \cgs}\\
\hline
X/O	&  Tot.     & Id.& compl. & Tot.      & Id. & compl. \\
\hline
$>10$	&  22       & 1  & 5\%  & 43  & 19 & 44\%  \\
$3-10$	&  29       & 7  & 24\% & 36  & 30 & 83\%  \\
$0.1-3$	&  72       & 54 & 75\% & 97  & 94 & 97\%  \\
\hline
\end{tabular}

\end{table*}

Notably, both above and below $10^{-14}$ \cgs the number of objects
with X/O$>3$ is comparable to that of those with X/O=0.1-3. But if we
consider the degree of completeness in the identifications, thanks to
the HELLAS2XMM 1dF survey, it is much higher above $10^{-14}$ \cgs,
than below.

At this point we could proceed immediately to estimate lower limits to
the number density of AGN, irrespective of their type, using only the
fraction of them with available redshifts. As just noted, however,
this fraction is not equally representative of the whole in different
bins of X/O and flux. As we shall show in the next Sect., a very
different behaviour of the X/O versus L$_{2-10keV}$ is found in AGN1 and in
the rest of the population. These behaviours give the opportunity to
statistically attribute a distance also to the sources without a
redshift determination, hence for the main purpose of this paper the
whole sample could be treated as if it were completely identified.

\subsection{Optically obscured vs. optically unobscured AGN}

Table 3 gives for the same bins of X/O and flux of Table 2 the ratio
between optically obscured to optically unobscured AGN.  Altough the
errors on the ratios are large, a trend toward a higher fraction of
optically obscured objects is suggested going both toward higher X/O
values and toward lower fluxes.  In order to confirm these trends one
should perform a wide ($\approx1$ deg$^2$) {\em and} deep
(F$_{2-10keV}>1-3\times10^{-15}$ \cgs~) X-ray survey. To extend the
redshift determinations to the sources with X/O=3-30 at this flux
limit (i.e. at R=25-27) deep optical to near infrared follow-ups are
needed, to obtain reliable photometric redshifts.  We also note that
Fig. \ref{hrt} suggests, at least qualitatively, the same kind of
trend, i.e. an average increase of obscuration (directly in the X-rays
band in this case) going toward low fluxes.
 
\setcounter{table}{2}    
\begin{table*}[ht]
\caption{\bf Ratio of optically obscured to optically unobscured AGN}
\begin{tabular}{lcccccc}
\hline
\hline
X/O	& F$_{2-10keV}<10^{-14}$ \cgs & F$_{2-10keV}>10^{-14}$ \cgs \\ 
\hline
$>10$	&  -                      & 3.8$^{+3.2}_{-2.0}$ \\
$3-10$	&  2.5$^{+3.7}_{-2.0}$   & 1.1$^{+0.5}_{-0.4}$ \\
$0.1-3$	&  1.21$^{+0.36}_{-0.30}$ & 0.31$^{+0.09}_{-0.07}$ \\
\hline
\end{tabular}

\end{table*}

\begin{figure*}[t]
\centering 
\includegraphics[angle=-90,width=16cm]{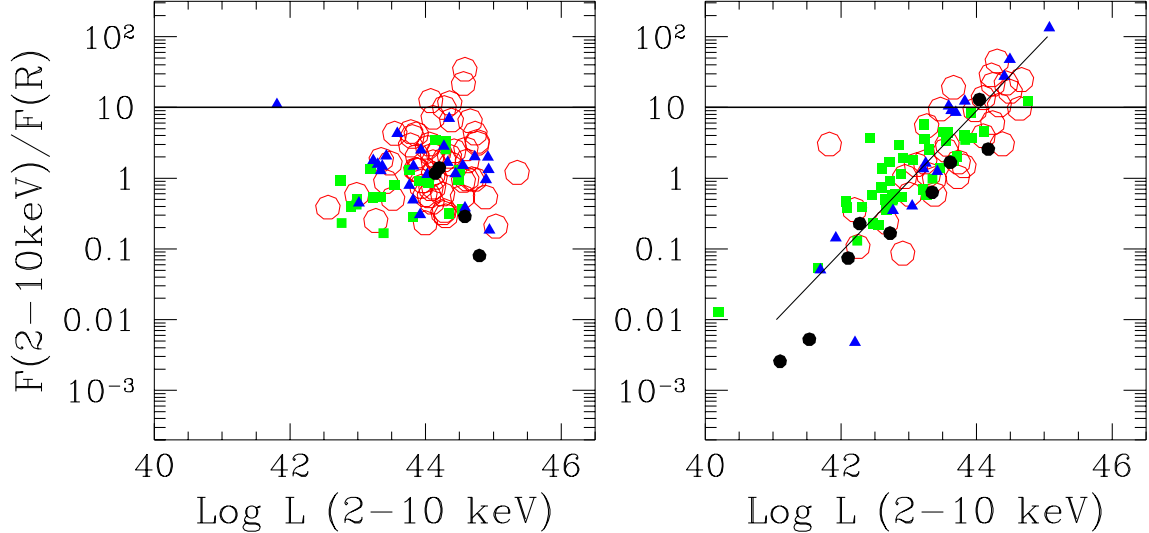}
\caption{The X-ray to optical flux ratio
as a function of the X-ray luminosity 
for type 1 AGN, left panel and non type 1 AGN and galaxies
(right panels). Symbols as in Fig. \ref{hrt}. The orizontal
lines mark the level of X/O=10, $\sim20\%$ of the sources in the
combined sample have X/O higher than this value. The diagonal line in the
right panel is the best linear regression between log(X/O) and
logL$_{2-10keV}$.
}
\label{xolx}
\end{figure*}

Fig. \ref{xolx} show the X-ray to optical flux ratio
as a function of the X-ray luminosity for broad line AGN (left panel)
and non broad line AGN and galaxies (right panel).

The average X/O for the 113 broad line AGN of the combined sample is
1.2 with a standard deviation of 0.3.  The average X/O of optically
selected QSO is 0.3, with a standard deviation of 0.4, using the ASCA
and BeppoSAX fluxes (George et al. 2000, Mineo et al. 2000) of 35 PG
QSO.  The difference in the average is due to the fact that optical
selection produces a tail toward low X/O: $\sim15\%$ of
the optically selected QSO are very X-ray faint, see Laor et
al. (1997), having X/O$<$0.1, against $\sim1\%$ for the broad line AGN
in the combined sample.

The right panel of Fig. \ref{xolx} shows the X-ray to optical flux
ratio as a function of the X-ray luminosity for optically obscured
AGN.  There is here a striking correlation between X/O and
L$_{2-10keV}$: higher luminosity AGN tend to have higher X/O.  The
solid diagonal line in the panel represents the best linear regression
between log(X/O) and logL$_{2-10keV}$ (a least square fit gives a slope
slightly flatter that this regression). A very similar correlation is
obtained computing the ratio between the X-ray and optical
luminosities, instead of fluxes (because the differences in the K
corrections for the X-ray and optical fluxes are small in comparison
to the large spread in X/O).

All objects plotted in the right panel of Fig. \ref{xolx} do not
show broad emission lines, i.e. the nuclear optical-UV light is
completely blocked, or strongly reduced in these objects, unlike the
X-ray light. Indeed, the optical R band light of these objects is
dominated by the host galaxy and therefore, {\em X/O is roughly a
ratio between the nuclear X-ray flux and the host galaxy starlight
flux}.  While the X-ray luminosity of these objects spans about 4
decades, the host galaxy R band luminosity has a moderate scatter, less
than one decade, around the mean value of $10^{11}$ L$_{\odot}$,
rather independent of redshift.

\subsection{Statistical predictions for the unidentified sources}

We can now use the fractions of obscured to unobscured objects in
Table 3 and the correlations in Fig. \ref{xolx} to predict the
luminosities, and therefore the redshifts, of the sources in the
combined sample without optical spectroscopic identification.  The
procedure is as follows: the sources without redshifts are divided in
the same bins of X/O and fluxes as in Tables 2 and 3 (the few lower
limits on X/O have been conservatively considered as measurements);
next they are picked randomly according to the ratios in Table 3 to
belong either to one or the other class.  For the lower flux, X/O$>10$
bin, for which we do not have information, we use the same ratio as in
the X/O$>10$, higher flux bin.  For those falling among the optically
obscured AGN we associate to each source a luminosity according to its
X/O value and the linear regression drawn in Fig. \ref{xolx} right
panel (we note that these luminosities are slightly lower than
those that would be obtained using the slightly flatter correlation
given by a least square fit, see the previous Sect.).
For those falling among the optically unobscured sources we
associate to each source a luminosity choosen at random from a
distribution similar to that of the broad line AGN in Fig.
\ref{xolx} left panel.  From the luminosities we derive redshifts,
using the same K correction and cosmology as for the sources with
spectroscopic (or photometric) redshift identifications.

Because the number of objects with measured redshift in the bins with
X/O$>3$ is relatively small and therefore the errors on the ratio
between obscured and unobscured objects are large, we tested several
different values within the error interval. The spread introduced in
the redshift and luminosity distributions by these uncertainty is
smaller than 10\%.

\begin{figure}
\centering 
\includegraphics[angle=0,width=9cm]{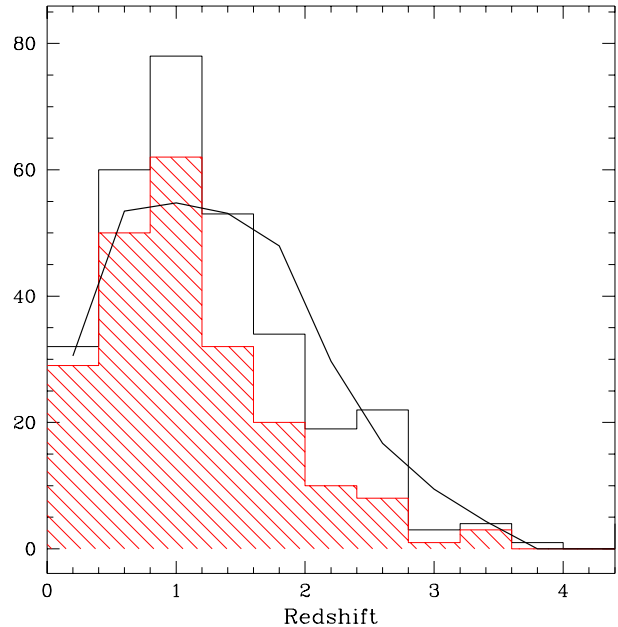}
\caption{ The z distribution of the 310 sources in the combined sample.
Shaded histogram = distribution of the spectroscopically identified
sources (70 \% of the sample).  Solid histogram = full sample.  
The luminosity and the redshift 
of the sources without a spectroscopic redshift have been
estimated using the correlation in Fig. \ref{xolx}, and the fraction
of obscured to unobscured sources in Table 3, see Sect. 5.2 for details.
The black thick line represents the expectation of the AGN synthesis
models of the CXB (Comastri et al. 2001), folded through the
appropriate sky-coverage, where unobscured and obscured AGN follow the
same pure luminosity evolution.  }
\label{zdist}
\end{figure}

\begin{figure}
\centering 
\includegraphics[angle=0,width=9cm]{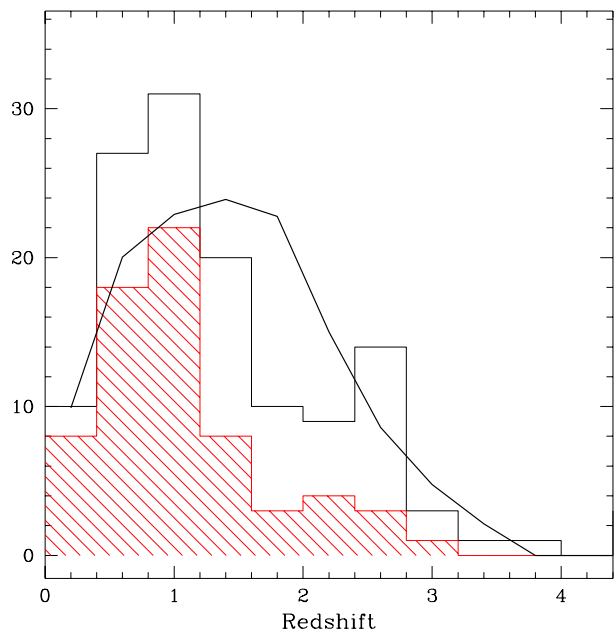}
\caption{ The z distribution of the 132 sources with
F$_{2-10keV}<10^{-14}$ \cgs. Shaded histogram = distribution of the
spectroscopically identified sources (53 \% of the sample).  Solid
histogram = full sample. See Fig. \ref{zdist} for details.
}
\label{zdist2}
\end{figure}

The z distribution of the 310 sources in the combined sample is
plotted in Fig. \ref{zdist} while the z distribution of the 132
sources in the combined sample with F$_{2-10keV}<10^{-14}$ \cgs is
plotted in Fig. \ref{zdist2}. The histograms of the sources with
measured redshift in Fig. \ref{zdist2} show a sharp decrease at
z=1.2 (similar results are presented also by Franceschini et al. 2002
and Hasinger 2003).  When the sources with an estimated redshift are
added, the peak of the redshift distribution at z$\sim1$ is confirmed,
but the decrease above this redshift is less sharp, thus suggesting
that this feature in the distribution of the sources with
spectroscopic redshift is enhanced by the incompleteness of the
optical identification (in particular at high X/O values, see Table
2).

\section{The evolution of hard X-ray selected sources}

We have used the combined sample of 310 hard X-ray selected sources,
plus 66 sources from the HEAO1 A2 all sky survey (Grossan et al. 1992)
with F$_{2-10keV}>2\times10^{-11}$ \cgs, to compute the evolution of
the number density and of the X-ray luminosity density. We have
adopted the standard 1/V$_{max}$ method (Schmidt 1968, Lilly et
al. 1995, Cowie et al. 2003). While it is well known that this method
is not free from biases (the main one is that it does not take into
account for evolution within each L and z bins), it is robust enough
to derive general trends (see e.g. Cowie et al. 2003), which is the
purpose of this paper. A more detailed computation of the evolution of
the luminosity function using a maximum likelihood method is deferred
to a future publication (La Franca et al. in preparation).

We are well aware that adopting a fixed ``average'' spectral index to
compute the K corrections is likely to introduce a bias in both the
luminosity distribution and the distribution of the maximum volume
allowed at each flux limit, which will affect the evaluation of the
luminosity function and its integrals (because the K correction is,
roughly, a power law function of the redshift). To evaluate the
magnitude of the systematic uncertainties on the luminosities and on
number and luminosity densities at high z due to this bias we
performed a number of dedicated simulations assuming a gaussian
distribution of spectral indices with $\sigma(\alpha_E)=0.2$, about
the value found for samples of AGN bright enough
to allow reliable spectral indices determinations through proper
spectral fittings (see e.g. Mineo et al. 2000, George et al. 2000).  
We find that the number and luminosity densities
at z$>1-2$ obtained assuming a fixed spectral index are higher than
those obtained assuming a gaussian distribution by 8-10\% at most.
This is smaller than both the statistical errors and the other
systematic errors affecting such computations (see Sect. 5.2).  In
the following we present the results obtained using a power law
spectrum with $\alpha_E$ fixed to 0.8 and reduced at low energy by
photoelectric absorption.

\begin{figure}
\centering
\includegraphics[angle=0,width=9cm]{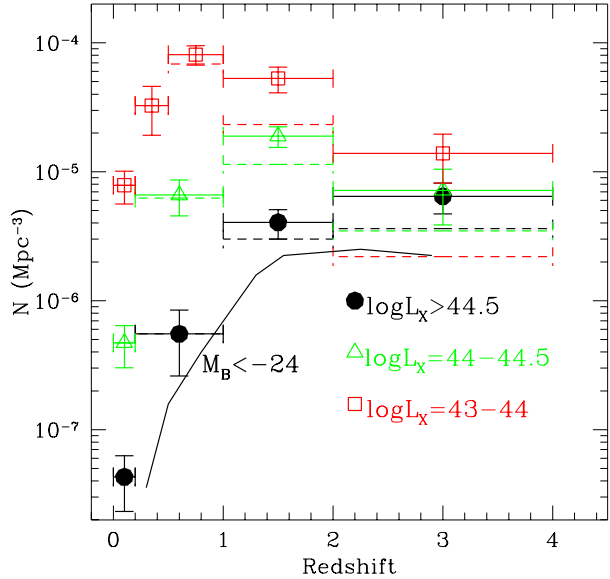}
\caption{ The evolution of the number density of hard X-ray selected
sources in three bins of luminosity: logL$_{2-10keV}$=43-44 erg
s$^{-1}$ = empty squares; logL$_{2-10keV}$=44-44.5 erg s$^{-1}$ =
empty triangles; logL$_{2-10keV}>44.5$ erg s$^{-1}$ = filled circles.
Note that the z=2--4, logL(2-10)=43--44 bin is a lower limit, see the
text for details.  Dashed lines represent lower limits obtained using
only the sources with measured redshift, see the text.  The solid
continuous curve represents the evolution of optically selected QSO
more luminous than M$_B=-24$. Note as the shape of the solid curve is
similar to the evolution of the luminous X-ray selected sources.  }
\label{evol}
\end{figure}

We have computed the evolution of both the number density of hard
X-ray sources and of the 2-10 keV luminosity density in three bins of
luminosities: 43$<$logL(2-10)$<$44 44$<$logL(2-10)$<$44.5 and
44.5$<$logL(2-10)$<$46.  Fig. \ref{evol} plots the evolution of the
number density for these three luminosity bins.  We emphasize that the
value for the z=2--4, logL(2-10)=43--44 bin is actually a lower limit,
because at our flux limit the objects with logL(2-10)$<$43.5 are not
accessible at z$>$2. In addition, if only the sources with a measured
z are considered, we obtain the lower limits plotted as dashed lines
in Fig.  8. We see that the number density of lower luminosity AGN
increases between z=0 and z=0.5 by a factor $\sim13$. It stays
constant up to z$\sim2$, while at higher z we cannot obtain a reliable
estimate of the number density behaviour for the reasons explained
above.  Conversely, the number density of luminous AGN increases by a
factor $\sim100$ up to z=2 and by a factor $\sim170$ up to z$\sim3$.
The last behaviour is similar to that of luminous (M$_B<-24$)
optically selected AGN (Hartwick \& Shade, 1990), solid thick line in
Fig. \ref{evol}.  The different evolution of low and high luminosity
sources is confirmed if we consider the sample of the identified
sources only.

\begin{figure}
\centering
\includegraphics[angle=0,width=9cm]{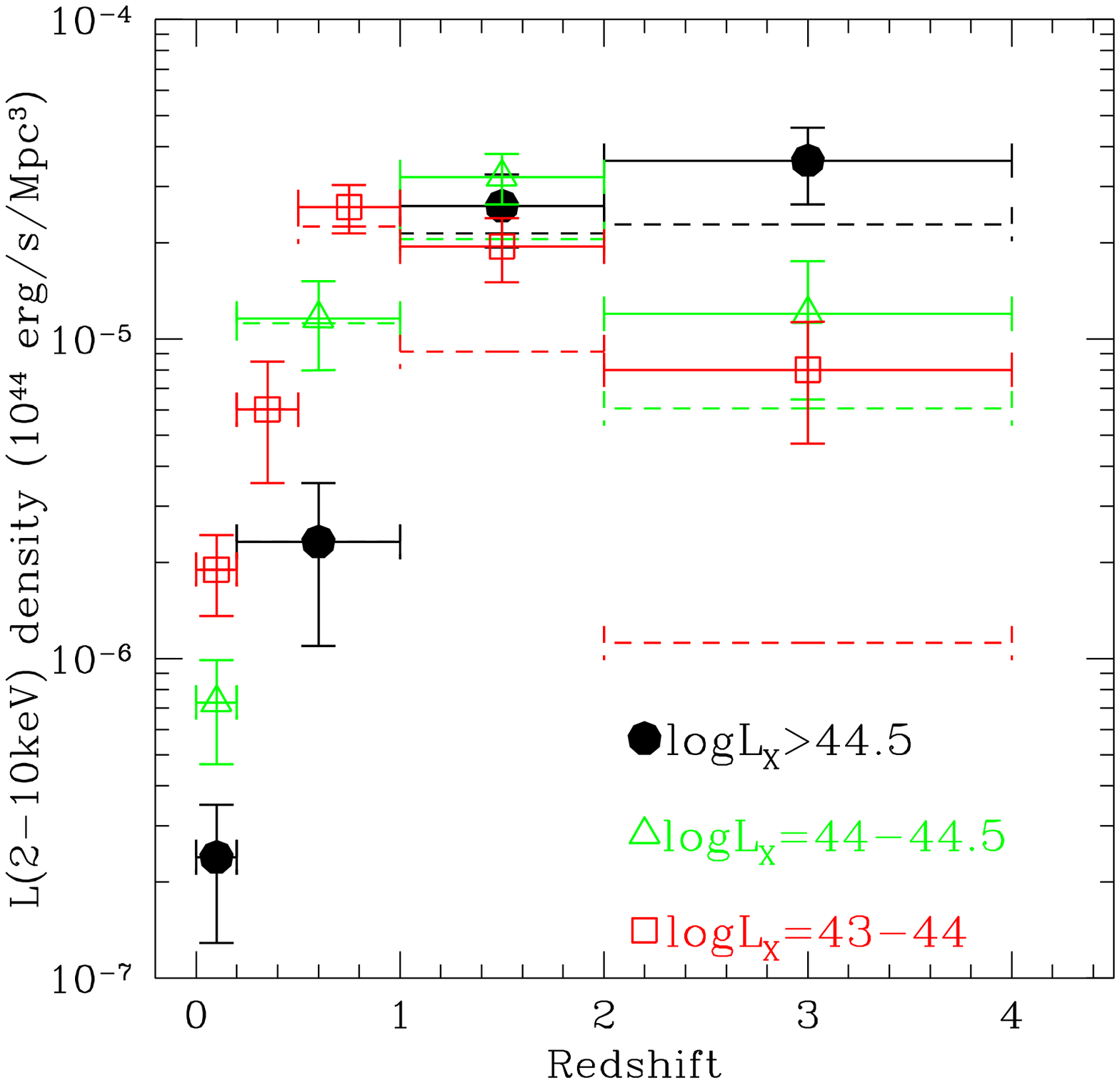}
\caption{ The evolution of the luminosity density of hard X-ray
selected sources in three bins of luminosity: logL$_{2-10keV}$=43-44
erg s$^{-1}$ = empty squares; logL$_{2-10keV}$=44-44.5 erg s$^{-1}$ =
empty triangles; logL$_{2-10keV}>44.5$ erg s$^{-1}$ = filled circles.
Note that the z=2-4, logL(2-10)=43-44 bin is a lower limit, see the
text for details. Dashed lines represent lower limits obtained using
only the sources with measured redshift, see the text.  }
\label{evoll}
\end{figure}

Fig. \ref{evoll} plots the evolution of the hard X-ray luminosity
density.  We see that the luminosity density of lower luminosity AGN
increases from z=0 to z$\sim1.5$ by a factor of $\sim18$, while that
of high luminosity AGN increases up to z=2 by a factor $\sim100$ and
up to z$\sim3$ by a factor $\sim170$.

\section{Discussion and conclusions}

We have obtained optical photometry and spectroscopy of a sample of
122 sources detected in the 2-10 keV band in five XMM-Newton fields of
the HELLAS2XMM serendipitous survey. The number and luminosity density
of AGN up to z$\sim3$ is derived by combining  ours with other
samples from deeper Chandra and XMM-Newton surveys (CDFN, SSA13,
Lockman Hole). The approach is rather similar to that followed by
Cowie et al (2003), based on a combination of the CDFN and a ROSAT
sample. There are, however, significant differences, hence, before
discussing the results, we wish to stress the major asset contributed
by our medium-deep survey.

About 20\% of the sources in our sample show an X-ray to optical flux
ratio ten or more times higher than typical of optically selected AGN.
A similar fraction of high X/O hard X-ray selected sources is also
present in deeper Chandra and XMM-Newton surveys (CDFN, CDFS, Lockman
Hole).  At the flux limit of the HELLAS2XMM 1dF sample several sources
with X/O$\gs10$ have optical magnitudes R=24-25, bright enough for
reliable spectroscopic redshifts to be obtained with 10m class
telescopes.  Indeed, we were able to obtain spectroscopic redshifts
and classification of 13 out of the 28 HELLAS2XMM 1dF sources with
X/O$>10$: the large majority of these sources are type 2 QSO,
i.e. high luminosity (logL$_{2-10keV}>44$), narrow line AGN at
z=0.7-1.8.

Conversely, at the $\sim10$ times lower fluxes probed by ultra-deep
Chandra and XMM-Newton surveys the optical magnitude of the sources
with X/O$\gs10$ is R$\gs27$, not amenable at present to optical
spectroscopy.  Indeed only 9 out of the 53 sources in the combined
sample with F$_{2-10keV}<10^{-14}$ \cgs and X/O$>3$ have spectroscopic
redshifts (and 1 among the 23 sources with X/O$>$10, see Sect. 5 and
Table 2). This limitation leads to a strong bias in ultra-deep Chandra
and XMM-Newton surveys against AGN highly obscured in the optical,
i.e. against type 2 QSO.  In fact, only 4 type 2 QSO have been
identified in the CDFN sample used in this paper (and 6 in the CDFS
sample, Hasinger 2003).

This bias renders the ultra-deep surveys alone inadequate to sample
properly the redshift distribution and luminosity functions. To
partially (or preliminarly) overcome this problem, we have used the
results obtained at higher X-ray (and optical) fluxes for the
HELLAS2XMM 1dF sample to derive, with a statistical approach, an
estimate of the luminosities, hence of the redshifts, of the sources
in the combined sample without optical spectroscopic
identification. The approach is based on: a) the fraction of optically
obscured to unobscured AGN found at F$_{2-10keV}>10^{-14}$ \cgs and
X/O$>3$; and b) the strong correlation found between the X-ray to
optical flux ratio and the X-ray luminosity in obscured AGN (see
Fig. 5b). It is worth noting that the X-ray to optical flux and
luminosity ratios of non broad line AGN might therefore be used to
investigate and constrain the relationship between the supermassive
black hole masses and efficiency of accretion, and bulge optical
luminosity (e.g. Gebhardt et al. 2000). This is beyond the scope of
this paper and will be matter of a follow-up publication.

The combined sample of 310 sources (70\% with robust spectroscopic
redshifts and 30\% with luminosities and redshifts assigned
statistically) was then used to derive the redshift distribution and
the cosmological evolution of the number and luminosity densities of
accretion dominated, hard X-ray selected sources.

The redshift distributions in Fig. \ref{zdist} and \ref{zdist2}
show a clear peak at z=1, but, when the objects with the "estimated"
redshifts are included, the drop above z=1.2 is rather smoother than
in the plots shown in Franceschini et al. (2002) and Hasinger (2003).
The solid thick curves in Fig. \ref{zdist} and \ref{zdist2}
represent the expectations from the baseline AGN synthesis models of
the CXB presented in Comastri et al. (2001), where unobscured and
obscured AGN, in a fixed proportion, share the same, pure luminosity
cosmological evolution.  Fig. \ref{zdist2} in particular shows that
this model predicts a much broader peak between z=1 and z=1.8, falling
short of the observed distribution at z$<1$ and demanding for a
slightly higher number of objects at z=1-2. The agreement is instead
rather good at z$>2$. The contrast does not appear as dramatic as,
e.g., in Franceschini et al. (2002) and Hasinger (2003), but is
telling us that at least one of the main assumptions, namely the pure
luminosity evolution, which tends to broaden the peak, is not
supported by the data, as definitely shown by the next step.

The evolution of high luminosity AGN (logL$_{2-10keV}>44.5$ erg
s$^{-1}$) in Fig. \ref{evol} and \ref{evoll} rises monotonically from
z=0 to z$\sim3$ (quickly from z=0 to z=1.5 and then more gradually
above this redshift). Their number and luminosity densities increase
by more than two orders of magnitudes, similar to those of the high
luminosity, optically selected QSO (e.g. Hartwick \& Shade 1990). By
contrast, the low luminosity AGN (logL$_{2-10keV}=43-44$ erg s$^{-1}$)
densities rise by a factor of 15-20 only, up to z$\sim0.7$, and stay
about constant up to z = 2.
The dashed lines in Fig. \ref{evol} and \ref{evoll} are lower limits
computed using only the sources with spectroscopic (or photometric in
three cases) redshifts. The limits are very close to the points up to
z=1, indicating that up to this redshift our determination of the
number and luminosity density of both high and low luminosity AGN is
very reliable.  For z$>$1 and logL$_{2-10keV}=43-44$ erg s$^{-1}$ the
limits are much lower than the points, which are mostly weighted by
our statistical estimates.  We note that points and lower limits in
each luminosity bin of Fig. \ref{evol} and \ref{evoll} are, by
construction, all {\em independent}, which make them easier to
compare, unlike the upper limits in Fig. 4 of Cowie et al. (2003).  If
our redhift estimates are correct, the number density in the
logL$_{2-10keV}=43-44$ erg s$^{-1}$ bin is maximal in the range
z=0.5-2 and stays higher than that of the sources with
logL$_{2-10keV}>44.5$ erg s$^{-1}$ up to z=3--4 (similarly to what is
found in soft X-ray selected AGN, Hasinger 2003). The luminosity
density at z$<1$ is dominated by low luminosity AGN, while (within the
large statistical error bars and the uncertainty associated with the
assumptions made in assigning a redshift when not directly available)
at z$>1$ those of low and high luminosity AGN are comparable, and
similar to that of low luminosity AGN at z=0.5--1.

All these results are qualitatively in agreement with those obtained
by Cowie et al. (2003). Quantitatively, we note that the integrated
luminosity density we derive is higher than the value given by Cowie
et al. (2003), although within their uncertainty.  We obtain for
logL$_{2-10keV}>43$ $4.6\times10^{39}$ erg s$^{-1}$ Mpc$^{-3}$ at
z=0.1--1 and $5.6\times10^{39}$ erg s$^{-1}$ Mpc$^{-3}$ at z=2--4,
against $\sim2\times10^{39}$ erg s$^{-1}$ Mpc$^{-3}$, with a factor of
$\sim3$ uncertainty, in Cowie at al. in both redshift bins.  This is
probably due to the different prescriptions in computing luminosities
(Cowie et al., 2003, do not correct for intrinsic absorption), and to
the fact that deep but small area surveys probe mostly the flatter,
low luminosity end of the AGN luminosity function at z$<1-2$, while
larger area surveys, like the HELLAS2XMM 1dF sample, push the AGN
selection to 3-10 times higher luminosities (at each given z$<1-2$,
see Fig. \ref{zlx}), up to and above L$^*$, which identifies the
break in the luminosity function, whereabouts most of the contribution
to its integral comes from. We note that following the same reasoning
of Cowie et al. (2003), and assuming a constant integrated luminosity
density evolution, we would find a present universal supermassive
black hole density of $4-5\times10^5$ M$_{\odot}$ Mpc$^{-3}$, about 2-3
times higher than that estimated by Cowie et al. (2003), but
consistent with both that measured from the intesity of the CXB
(Fabian et al. 1999, Elvis et al. 2002, Fabian 2003) and from local
galaxies (Gebhardt et al. 2000, Ferrarese \& Merrit 2000).

The qualitative picture emerging from both observational data and
theories of galaxy and AGN evolution is intriguing.  The evolution of
luminous AGN is strong up to at least z=2-3 and, as suggested by
e.g. Franceschini et al. (1999) and Granato et al. (2001), may follow
the evolution of spheroids.  Powerful AGN are likely to be present in
these massive galaxies.  Furthermore, the radiation and the strong
winds produced by these active nuclei may help inhibiting the
star-formation in these galaxies, which therefore would have red
colors. Intriguingly, Mignoli et al. (2003) find a strong correlation
between the R-K color and the X/O ratio for a sample of HELLAS2XMM 1dF
sources.  Moreover, all the 10 sources in the Mignoli et al. sample
with X/O$>$10 have R-K$\gs5$, i.e. they are all Extremely Red Objects.

On the other hand, the number and luminosity densities of lower
luminosity AGN evolve up to z=1 and possibly decrease at z$>$2. A
similar behaviour is observed in galaxies in which the star-formation
is ongoing (Lilly et al. 1995, Madau et al. 1996, 1998, Fontana et
al. 1999).  A link between the formation and evolution of galaxies and
the growth and light-up of supermassive black holes in their nuclei
has been investigated in more detail by Cavaliere \& Vittorini (2000)
and by Menci et al.  (2003) in the framework of bottom-up models: the
latter authors were able to reproduce rather naturally the steep
evolution which is observed in the highly luminous quasar
population. Menci et al. (2003) compare their model with the
luminosity function and the number densities of luminous optically
selected quasars only (since little is know about the evolution of low
luminosity, optically selected AGN, because standard color techniques
cannot be used when the nucleus luminosity is similar or smaller than
the host galaxy luminosity). The results presented in this paper about
the differential evolution of high and low luminosity X-ray selected
AGN can therefore provide additional and more tight contraints to
hierarchical clustering models. This will be matter of a future
publication (Menci et al. in preparation).

It is clear that while the general trends are rather robust (several
independent analyses converge to the similar qualitative results) the
number of X-ray sources with identified redshift is still far too
small to derive accurate quantitative information about the
differential AGN evolution at z=2-4. Furthermore, the sample sizes are
insufficient to attempt deriving the evolution of obscured AGN
separately from that of unobscured AGN, so that key questions such
as whether they are similar or not, and whether the amount of
obscuration is or not related to the luminosity remain open.  The
above limitations can be overcome by increasing the area covered in
X-ray surveys and their optical-NIR photometric and spectroscopic
follow-up, both at relatively high (F$_{2-10keV}>10^{-14}$ \cgs) and
low (F$_{2-10keV}=1-5\times10^{-15}$ \cgs) fluxes.  For example, in
the HELLAS2XMM 1dF sample there are some thirty "optically obscured"
AGN: 100-150 sources of this type would be sufficient to adequately
figure their luminosity function over 2-3 luminosity dex and a few
redshift bins. This is the goal of the extension of the optical
follow-up of the HELLAS2XMM survey from 1 to 4 square degree, which
should allow us to contrast the luminosity function of obscured and
unobscured AGN, and to study their differential evolution up to z$\sim$2.
On the other hand, the ultra-deep but small area CDFN survey has
provided so far only 6 AGN more luminous that logL$_{2-10keV}=44$ at
z$>3$ and 20 at z$>2$ (Cowie et al. 2003).  Note that all of them have
flux $\gs10^{-15}$ \cgs, suggesting that the most effective strategy
to find high luminosity, high z AGN consists in increasing the area
covered at F$_{2-10keV}=1-5\times10^{-15}$ \cgs, rather than pushing
the depth of the survey. Observing $\sim0.5$ square degrees of sky
(roughly 10 times the area covered by the CDFN survey) at the above
flux limit, would roughly increase by a factor of 10 the number of
high z objects, a feasible program with a reasonable amount of
observing time.  The related issue of obtaining redhifts is more
complex, because the optical counterparts of a large fraction of them
will be too faint for optical spectroscopy. The alternative will be
accurate multiband optical to near infrared photometry, in order to
obtain reliable photometric redshifts through the comparison of the
observed Spectral Energy Distribution with templates of galaxy and AGN
spectra.

\setcounter{table}{0}    
\begin{table*}[ht]
\caption{\bf The HELLAS2XMM 1dF sample}
\begin{tabular}{lcccccccccc}
\hline
\hline
Id & X-ray Ra & X-ray Dec & optical RA & Optical Dec & Diff. &
F$_{2-10keV}$ & R & Class. & z & logL$_{2-10keV}$ \\
   &  2000    & 2000      & 2000       & 2000        & arcsec &
$10^{-14}$ cgs & &     &    & erg/s \\       
\hline
05370037$^a$&05 41  00.4 & -28 39  05 &05 41  00.7 & -28 39  01 & 5.8  &  4.93 & 21.5 & -- &  --    & -- \\
05370015 &05 40 54.3 & -28 43  45 &05 40 54.2 & -28 43  47 & 2.4    &  3.19 & 19.9 &AGN1&  0.880 & 44.03 \\
05370022$^{b}$ &05 40 51.3 & -28 36  42 &05 41 51.4 & -28 36 46&4.3&2.40 & $\gs$23.0 & -- &  --  & -- \\
05370175 &05 40 45.6 & -28 39   07 &05 40 45.7 & -28 39   08 & 1.7  &  2.08 & 19.8 &AGN1&  1.246 & 44.32 \\
05370008 &05 40 34.2 & -28 31   08 &05 40 34.2 & -28 31   09 & 1.2  &  5.63 & 19 &ELG&  0.379 & 43.44 \\  
05370021 &05 40 26.1 & -28 50  39 &05 40 26.3 & -28 50  42 & 4.2    &  3.22 & 23.5 &ELG&  1.192 & 44.35 \\
05370135 &05 40 24.7 & -28 46  16 &05 40 24.6 & -28 46  16 & 1.3    &  1.21 & 21.3 &AGN2&  0.484 & 43.02 \\
05370043 &05 40 22.1 & -28 31  40 &05 40 22.1 & -28 31  40 & 0.2    &  3.06 & 22.7 &AGN2&  1.797 & 44.77 \\
05370091 &05 40 21.2 & -28 50  38 & 05 40 21.1 & -28 50 38  & 1.2& 2.50 & 23.7 & -- &  --  & --  \\
05370162 &05 40 13.0 & -28 44   02 &05 40 13.0 & -28 44   02 & 0.0  &  0.86 & 21.6 & -- &  --    & --  \\
05370019 &05 40 10.6 & -28 40  51 &05 40 10.5 & -28 40  53 & 2.7    & 1.62 & 20.3 &AGN1&  1.330 & 44.15 \\
05370072 &05 40  04.4 & -28 38  14 & --        &  --      &   --    &  0.83 & $\gs$24 & -- &  --    & -- \\
05370159 &05 40  02.7 & -28 37  29 & --       & --       &   --     &  1.06 & $\gs$24 & -- &  --    & -- \\
05370031 &05 40  00.8 & -28 34  56 &05 40  00.8 & -28 34  56 & 0.2  &  1.30 & 20.5 &AGN1&  3.276 & 44.96\\
05370007 &05 39 59.0 & -28 37  53 &05 39 59.0 & -28 37  53 & 0.2    &  2.90 & 20.6 &AGN1&  0.842 & 43.93 \\
05370035 &05 39 58.7 & -28 37   07 &05 39 58.7 & -28 37   08 & 1.4  &  1.24 & 22.8 &AGN1&  0.897 & 43.65\\
05370157 &05 39 58.6 & -28 41  26 &  --      & --       &   --      &  0.77 & $\gs$24.5 & -- &  --  & --\\
0537042a &05 39 57.5 & -28 49  14 &05 39 57.2 & -28 49  14 & 3.8    &  1.39 & 21.6 &AGN1&  1.945 & 44.48 \\
0537042b &05 39 58.8 & -28 49  19 &05 39 59.0 & -28 49  19 & 2.0    &  2.10 & 21.5 & -- &  --    & --    \\
05370013 &05 39 57.2 & -28 51   09 &05 39 57.0 & -28 51   08 & 2.6  &  2.54 & 22 &AGN1&  0.901 & 43.96 \\  
05370017 &05 39 57.0 & -28 50  28 &05 39 56.7 & -28 50  27 & 3.9    &  2.62 & 20.7 &AGN1&  0.904 & 43.96 \\
05370153 &05 39 56.1 & -28 46  22 &   --     & --       &   --      &  1.13 & $\gs$24.6 & -- &  --  & --\\
05370012 &05 39 52.7 & -28 47   09 &05 39 52.8 & -28 47   09 & 1.5  &  1.49 & 22.5 & -- &  --    & --  \\
0537052a &05 39 50.4 & -28 33  45 &05 39 50.4 & -28 33  45 & 0.4    &  1.11 & 21.5 &AGN1&  1.665 & 44.24 \\
0537052b &05 39 51.9 & -28 33  45 &05 39 51.8 & -28 33  45 & 1.7    &  1.36 & 23.7 & -- &  --    & --    \\
05370004 &05 39 49.9 & -28 38  32 &05 39 50.0 & -28 38  32 & 1.5    &  7.02 & 18.0 &AGN1&  0.894 & 44.39 \\
05370054$^c$ &05 39 45.3 & -28 49  10 &05 39 45.2 & -28 49  10 & 1.3&  1.62 & 25.0 & -- &  --    & --  \\  
05370014 &05 39 43.0 & -28 27  20 &05 39 43.1 & -28 27  19 & 2.0    &  3.00 & 19.6 &AGN1&  1.659 & 44.68 \\
05370041 &05 39 39.7 & -28 31  43 &05 39 39.8 & -28 31  42 & 1.9    &  1.56 & 20.4 &AGN1&  1.644 & 44.38 \\
05370078 &05 39 38.7 & -28 48  07 &05 39 38.7 & -28 48  08 & 1.3    &  1.44 & 22.0 &AGN1&  1.622 & 44.33 \\ 
05370036$^d$ &05 39 38.7 & -28 52  49 &05 39 39.2 & -28 52  51 & 6.9&  5.65 & 20.6 &AGN1&  1.329 & 44.71 \\
05370006 &05 39 34.8 & -28 41  16 &05 39 34.8 & -28 41  16 & 0.1    &  2.02 & 16   & *  &--      & -- \\  
05370060 &05 39 33.6 & -28 35  19 &05 39 33.7 & -28 35  20 & 2.0    &  1.72 & 23.9 & -- &  --    & --   \\
05370003 &05 39 29.5 & -28 49  00 &05 39 29.4 & -28 49  00 & 1.2    & 11.21 & 18.1 &AGN1&  0.317 & 43.55 \\
05370024 &05 39 25.8 & -28 44  56 &05 39 25.7 & -28 44  54 & 2.0    &  4.57 & 21 &ETG &  0.075 & 41.82 \\  
05370010 &05 39 25.3 & -28 32  37 &05 39 25.4 & -28 32  36 & 1.5    &  4.24 & 22.4 & -- &  --    & --    \\
05370002 &05 39 23.5 & -28 42  24 &05 39 23.5 & -28 42  23 & 0.9    & 18.21 & 16.6 &AGN1&  1.244 & 45.14 \\
0537011a &05 39 20.5 & -28 37  22 &05 39 20.5 & -28 37  21 & 0.9    &  4.76 & 23.4 &AGN2$^e$&0.981& 44.33 \\
0537011b &05 39 21.6 & -28 38   04 &05 39 21.6 & -28 38   06 & 2.5  &  0.88 & 21.7 & -- &  --    & --    \\
05370040 &05 39 20.1 & -28 36  37 &05 39 20.1 & -28 36  38 & 1.1    &  1.54 & 21 &AGN1&  1.485 & 44.25 \\ 
05370164 &05 39 17.2 & -28 38  20 &05 39 17.1 & -28 38  17 & 3.3    &  1.49 & 23.6 &AGN2&  1.824 & 44.48 \\
05370111$^c$&05 39 11.5 & -28 37  17 &05 39 11.6 & -28 37  15 & 2.4 &  3.69 & 24.5 & -- &  --  &  --  \\
05370009 &05 39 10.7 & -28 35  28 &05 39 10.8 & -28 35  26 & 2.5    &  3.75 & 20.8 &AGN1&  0.770 & 43.95 \\
05370016 &05 39  09.3 & -28 41  05 &05 39  09.4 & -28 41   45 & 1.4 &  4.84 & 21.7 &AGN2$^f$&0.995 & 44.35 \\
05370005 &05 39  05.5 & -28 33  17 &05 39  05.5 & -28 33  16 & 1.2  &  8.94 & 21.1 &AGN1&  1.158 & 44.75 \\
05370123 &05 38 51.4 & -28 39  49 &05 38 51.5 & -28 39  49 & 1.4    &  7.53 & 23.1 &AGN1&  1.153 & 44.72 \\
05370020 &05 38 50.9 & -28 37  57 &05 38 50.9 & -28 37  58 & 0.8    &  5.43 & 20.7 &AGN1& 0.763 & 44.21 \\
\hline
\end{tabular}

\end{table*}

\setcounter{table}{0}    
\begin{table*}[ht]
\caption{\bf The HELLAS2XMM 1dF sample, continue}
\begin{tabular}{lcccccccccc}
\hline
\hline
Id & X-ray Ra & X-ray Dec & optical RA & Optical Dec & Diff. &
F(2-10keV) & R & Class. & z & logL(2-10keV) \\
   &  2000    & 2000      & 2000       & 2000        & arcsec &
$10^{-14}$ cgs & &     &    & erg/s \\       
\hline
03120012 &03 15 29.4 & -76 53  42 &03 15 28.8 & -76 53  40 & 2.5    &  2.78 & 21 &AGN1&  0.507 & 43.49 \\  
03120020 &03 14 16.8 & -76 56  00 &03 14 16.5 & -76 56   02 & 2.4   &  1.61 & 21.5 &ELG$^g$& 0.964 & 43.87 \\
03120010 &03 14 16.4 & -76 45  37 &03 14 16.5 & -76 45  36 & 0.7    &  2.13 & 19.6 &AGN1&  0.246 & 42.74 \\
03120022 &03 13 49.0 & -76 45  59 &03 13 49.3 & -76 46  00 & 1.3    &  2.71 & 21.6 &AGN1&  2.140 & 44.85 \\
03120036$^c$&03 13 43.5 & -76 54  26 &03 13 42.9 & -76 54  24 & 3.2 &  1.45 & 24.6 & -- &  --  & -- \\
03120013 &03 13 34.3 & -76 48  30 &03 13 34.5 & -76 48  30 & 0.9    &  1.17 & 19.7 &AGN1&  1.446 & 44.25 \\
03120003 &03 13 14.7 & -76 55  56 &03 13 14.3 & -76 55  56 & 1.2    & 16.70 & 18.3 &AGN1&  0.420 & 44.03 \\
03120005 &03 13 12.0 & -76 54  30 &03 13 11.8 & -76 54  31 & 1.4    &  8.37 & 19.1 &AGN1&  1.274 & 44.50 \\
03120127 &03 12 58.0 & -76 51  20 &03 12 57.9 & -76 51  20 & 0.4    &  1.00 & 23.5 &AGN1&  2.251 & 44.79 \\
03120065$^b$&03 12 52.2 & -77 00 59 & --     &  --      &   --      &  1.47 &$\gs$24& -- &  --   & --  \\
03120006 &03 12 54.0 & -76 54  15 &03 12 53.8 & -76 54  15 & 0.7    &  8.28 & 22 &AGN2&  0.680 & 44.16 \\  
03120018 &03 12 39.3 & -76 51  33 &03 12 38.7 & -76 51  33 & 1.9    &  2.59 & 18 &ETG &  0.159 & 42.24 \\  
03120008 &03 12 31.2 & -76 43  24 &03 12 31.3 & -76 43  25 & 1.1    &  1.64 & 13.7 &ETG$^h$& 0.052 & 41.08\\
03120004 &03 12  09.2 & -76 52  13 &03 12  09.2 & -76 52  13 & 0.1  &  6.49 & 18.2 &AGN1&  0.890 & 44.49 \\
03120016 &03 12  00.4 & -77 00  26 &03 12  00.5 & -77 00  26 & 0.6  &  1.51 & 22.2 &ELG&  0.841 & 43.79 \\
0312089a &03 11 45.0 & -76 56  45 &03 11 44.4 & -76 56  46 & 2.3    &  1.30 & 23.6 &ELG&  0.809 & 43.56 \\
03120181 &03 11 36.0 & -76 55  56 &03 11 36.0 & -76 56  00 & 3.8    &  1.10 & 23.2 &ELG&  0.709 & 43.49 \\
03120035$^i$&03 11 31.8 & -77 00  36 &03 11 31.8 & -77 00  32 & 4.3 &  1.83 & 22.0 &AGN1& 1.272 & 44.16 \\ 
03120066 &03 11 28.2 & -76 45  16 &03 11 29.2 & -76 45  15 & 3.7    &  1.51 & 23.1 &AGN1$^e$&1.449 &44.23\\
03120017 &03 11 24.8 & -77 01  39 &03 11 25.4 & -77 01  35 & 4.6    &  2.74 & 17.7 &ETG &  0.320 & 42.93 \\
03120031 &03 11 13.9 & -76 53  59 & 03 11 13.6 & -76 54  00 & 1.0   &  1.20 & 23.6 & -- & --  & -- \\
03120029 &03 11 13.4 & -76 54  31 &03 11 13.0 & -76 54  34 & 3.1    &  1.21 & 18.8 & -- &  --   & -- \\   
03120011 &03 11 12.8 & -76 47  02 &03 11 13.3 & -76 47  02 & 1.9    &  1.56 & 21.5 &AGN1&  0.753 & 43.72 \\
03120009 &03 11  05.6 & -76 51  58 &03 11  05.3 & -76 51  58 & 0.9  &  1.98 & 23.2 &AGN1&  1.522 & 44.45 \\
03120007 &03 10 50.0 & -76 39  04 &03 10 50.2 & -76 39  04 & 0.8    & 29.20 & 18.6 &AGN1&  0.381 & 44.12 \\
03120021 &03 10 49.8 & -76 53  17 &03 10 49.7 & -76 53  16 & 0.7    &  1.51 & 22.3 &AGN1&  2.736 & 44.85 \\
03120028 &03 10 37.4 & -76 47  13 &03 10 37.9 & -76 47  11 & 2.5    &  1.78 & 20.8 &ELG&  0.641 & 43.43 \\
03120045$^c$&03 10 19.0 & -76 59  58 &03 10 18.9 & -76 59  58 & 0.5 &  1.94 & 24.4 & -- &  --  & --  \\  
03120002 &03 10 15.8 & -76 51  33 &03 10 15.9 & -76 51  33 & 0.5    & 41.70 & 17.6 &AGN1&  1.187 & 45.50 \\
03120124 &03 10  01.6 & -76 51   07 &03 10  01.7 & -76 51   08 & 1.3 &  1.41 & 22.5 & -- &  --    & --    \\
03120501 &03 09 52.2 & -76 49  27 &03 09 51.0 & -76 49  23 & 5.9     &  1.36 & 20 &ETG &  0.205 & 42.22 \\  
03120014 &03 09 51.2 & -76 58  25 &03 09 51.4 & -76 58  26 & 1.5     &  3.97 & 18.4 &ELG&  0.206 & 42.66 \\
03120024 &03 09 31.7 & -76 48  45 &03 09 32.1 & -76 48  46 & 1.8     &  1.76 & 21.8 &AGN1&  1.838 & 44.58 \\
03120116 &03 09 18.5 & -76 57  59 &03 09 18.2 & -76 58  00 & 1.1    &  2.01 & 23.9 &ELG&0.814$^l$& 43.87 \\
03120034 &03 09 12.1 & -76 58  26 &03 09 12.1 & -76 58  26 & 0.2    &  9.80 & 19.1 &AGN2&  0.265 & 43.34 \\
\hline
\end{tabular}

\end{table*}

\setcounter{table}{0}    
\begin{table*}[ht]
\caption{\bf The HELLAS2XMM 1dF sample, continue}
\begin{tabular}{lcccccccccc}
\hline
\hline
Id & X-ray Ra & X-ray Dec & optical RA & Optical Dec & Diff. &
F(2-10keV) & R & Class. & z & logL(2-10keV) \\
   &  2000    & 2000      & 2000       & 2000        & arcsec &
$10^{-14}$ cgs & &     &    & erg/s \\       
\hline
26900038 & 23 59 57.2 & -25 05  43 & 23 59 57.2 & -25 05  44 & 1.4  &  2.25 & 21 &ELG&  0.904 & 43.93 \\  
26900075$^c$& 23 59 56.6 & -25 10  20 &  23 59 56.4 & -25 10 18 &3.3&  2.06 & 24.6 & -- &  --  & -- \\
26900039 & 23 59 39.8 & -25 00  57 & 23 59 39.8 & -25 00  57 & 0.2  &  3.90 & 19.6 &AGN1&  0.930 & 44.20 \\
26900028 & 23 59 33.4 & -25 07  58 & 23 59 33.3 & -25 07  57 & 2.0  &  3.30 & 21.8 &AGN1&  0.738 & 43.88 \\
26900006 & 00 01 22.8 & -25 00 19 & 00 01 22.7 & -25 00  19 & 1.4   & 12.20 & 18.7 &AGN1&  0.964 & 44.68 \\
26900029$^c$& 00 01 11.6 & -25 12  03 & 00 01 11.5 & -25 12  6& 3.9 &  3.56 & 25.1 & -- &  --  & -- \\
26900010 & 00 01  06.8 & -25 08 46 & 00 01  06.8 & -25 08  47 & 1.4 &  3.08 & 21 &AGN1&  1.355 & 44.47 \\  
26900003 & 00 01  02.4 & -24 58  47 & 00 01 02.4 & -24 58  49 & 1.7 & 11.16 & 20.3 &AGN1&  0.433 & 43.85 \\
26900002 & 00 01  00.2 & -25 04 59 & 00 00 59.9 & -25 05   00 & 4.2 & 14.35 & 21.9 &AGN1&  0.850 & 44.66 \\
26900014 & 00 00 44.3 & -25 07 38 & 00 00 44.3 & -25 07  40 & 2.2   &  1.63 & 21.6 & -- &  --    & --    \\
26900022 & 00 00 36.6 & -25 01  05 & 00 00 36.6 & -25 01   06 & 1.0 &  3.14 & 21.2 &AGN2&  0.592 & 43.64 \\
26900007 & 00 00 34.6 & -25 06 19 & 00 00 34.6 & -25 06  21 & 1.7   &  1.68 & 20.3 &AGN1&  1.234 & 44.22 \\
26900004 & 00 00 31.7 & -24 54  59 & 00 00 31.9 & -24 54  57 & 3.6  &  7.96 & 17.7 &AGN1&  0.284 & 43.35 \\
26900013 & 00 00 30.1 & -25 12  14 & 00 00 29.8 & -25 12  17 & 4.9  & 1.63 & 17.5 &ETG$^h$ & 0.154 & 42.07\\
26900001 & 00 00 27.7 & -25 04 41 & 00 00 27.7 & -25 04  43 & 1.6   &  7.81 & 19.1 &AGN1&  0.336 & 43.43 \\
26900012 & 00 00 26.0 & -25 06 48 & 00 00 26.0 & -25 06  51 & 2.5   &  1.70 & 20.3 &AGN1&  0.433 & 43.03 \\
26900015 & 00 00 22.8 & -25 12  20 & 00 00 22.9 & -25 12  22 & 2.2  &  1.74 & 19.7 &AGN1&  1.610 & 44.40 \\
26900009 & 00 00 21.2 & -25 08  13 & 00 00 21.2 & -25 08  12 & 1.4  &  2.16 & 20.9 &AGN1&  0.995 & 44.00 \\
26900072 & 00 00 13.7 & -25 20  11 & 00 00 13.6 & -25 20  13 & 2.1  &  4.05 & 23 &ELG&1.389$^l$ & 45.37 \\  
26900016 & 00 00  02.8 & -25 11  38 & 00 00 02.7 & -25 11  37 & 1.5 &  2.75 & 21 &AGN1&  1.314 & 44.37 \\  
15800002 & 00 34 19.1 & -11 59  37 & 00 34 19.0 & -11 59  39 & 2.3  &  9.64 & 20.2 &AGN1&  0.848 & 44.49 \\
15800012 & 00 34 18.5 & -12 08   09 & 00 34 18.5 & -12 08   09 & 0.5 &  6.79 & 18.8 &AGN2&  0.233 & 43.04 \\
15800011 & 00 34 15.4 & -12 08  47 & 00 34 15.5 & -12 08  45 & 2.9   &  1.84 & 20.7 &AGN1&  2.069 & 44.69 \\
15800062 & 00 34 13.9 & -11 56  00 & 00 34 13.8 & -11 56   00 & 1.1  &  4.49 & 23.3 &AGN2&  1.568 & 44.80 \\
15800025 & 00 34 10.0 & -12 11  26 & 00 34  09.9 & -12 11  32 & 5.8  &  2.56 & 21 &ELG&  0.470 & 43.37 \\  
15800019 & 00 33 57.3 & -12 00  40 & 00 33 57.2 & -12 00  38 & 1.9   &  3.12 & 21.8 &AGN2$^e$&1.957& 44.84\\
15800008 & 00 33 47.5 & -12 03  25 & 00 33 47.5 & -12 03  26 & 1.5   &  2.58 & 21 &AGN1&  1.151 & 44.22 \\  
15800092 & 00 33 42.5 & -12 01  34 & 00 33 42.7 & -12 01  37 & 4.6   &  2.65 & 22.8 &ELG$^e$&0.993 & 44.16\\
15800013 & 00 33 39.6 & -12 08  26 & 00 33 39.6 & -12 08  28 & 2.0   &  1.82 & 22.3 &ELG&1.326$^l$& 44.20 \\
15800005 & 00 33 24.1 & -12 06  51 & 00 33 24.1 & -12 06  49 & 2.5   &  2.35 & 20.2 &AGN1&  1.207 & 44.30 \\
15800017 & 00 33 20.6 & -12 05  40 & 00 33 20.6 & -12 05  38 & 1.3   &  2.83 & 20.3 &AGN1&  1.946 & 44.79 \\
15800001 & 00 33 15.7 & -12 06  57 & 00 33 15.7 & -12 06 59 & 1.8    &  7.29 & 19.9 &AGN1&  1.211 & 44.79 \\
50900036 & 20 44 46.4 & -10 38  40 & 20 44 46.7 & -10 38  43 & 5.4  &  3.30 & 20.2 &AGN2&  0.694 & 43.84 \\
50900001 & 20 44 28.7 & -10 56  29 & 20 44 28.4 & -10 56  33 & 5.6  &  4.97 & 23.9 &AGN2$^e$&1.049& 44.40 \\
50900061 & 20 44 20.5 & -10 49   04 & 20 44 20.5 & -10 49   03 & 1.6 & 4.22 & 20.4 &ETG &0.324$^l$ & 43.16\\
50900020 & 20 44 19.4 & -10 56  20 & 20 44 19.5 & -10 56  19 & 1.6  &  7.58 & 20.2 &AGN1&  0.770 & 44.30 \\
50900067 & 20 44  07.1 & -10 56  11 & 20 44  07.2 & -10 56  12 & 1.5 &  5.29 & 22.3 &AGN1&  1.076 & 44.44\\
50900013 & 20 43 49.7 & -10 32  44 & 20 43 49.5 & -10 32  40 & 5.0  &  4.42 & 23.2 &AGN2&  1.261 & 44.55 \\
50900031 & 20 43 49.2 & -10 37  46 & 20 43 49.1 & -10 37  43 & 3.0  &  6.13 & 20.6 &AGN1&  0.556 & 43.82 \\
50900039 & 20 43 22.8 & -10 40  30 & 20 43 22.5 & -10 40  31 & 4.7  &  5.19 & 22.4 &AGN1$^e$&0.818& 44.17\\
\hline
\end{tabular}

Classification: AGN1; AGN2; ELG=Emission Line Galaxy; ETG=Early Type Galaxy; 
*=Star; 
$^a$ Group of galaxies in the error-box; 
$^b$ Near bright star; 
$^c$ Aperture photometry at the position of a bright K source 
(Mignoli et al. 2003)
$^d$ Two possible counterparts;
$^e$ Classification tentative;
$^f$ Two nearby objects, a type 2 AGN and an emission line galaxy, 
at the same redshift;
$^g$ Two nearby emission line galaxies at the same redshift;
$^h$ Extended, emission from a group or cluster of galaxies, z based on nearest galaxy; 
$^i$ Two possible counterparts, a type 1 AGN and an R-K$\gs6.1$ object 
(Brusa et al. 2003, Mignoli et al. 2003);
$^l$ redshift tentative, based on a single faint line.

\end{table*}

%%%%%%%%%%%%%%
\end{document}